# The impact on health system expenditure in Australia and OECD countries from accelerated NCD mortality decline through prevention or treatment strategies to achieve Sustainable Development Goal Target 3.4

Bibha Dhungel[1], DrPH; Jingjing Yang[1], MSc; Tim Wilson[1], PhD; Samantha Grimshaw[1], MSc; Emily Bourke[1], MHE; Stephanie Khuu[1], PhD; Tony Blakely[1], PhD

[1]Population Interventions Unit, Melbourne School of Population and Global Health, The University of Melbourne, Carlton, VIC 3053, Australia

**\*Correspondence to**
Bibha Dhungel
Population Interventions Unit, The University of Melbourne
207 Bouverie St, Carlton VIC 3053
bibha.dhungel@unimelb.edu.au
+61 3 9035 5973

Total word count: 3460

**Abstract**

**Background**

It is unclear what the relative impacts of prevention or treatment of non-communicable diseases (NCDs) are on future health system expenditure. First, we estimated expenditure in Australia for prevention versus treatment pathways to achieve target 3·4 of the Sustainable Development Goals (SDG3·4). Second, we applied the method to all 34 other OECD countries.

**Methods**

Global burden of disease data was used to estimate average annual percentage changes (APCs) in disease incidence, remission and case fatality rates (CFRs) from 1990 to 2021 and projected to 2030 to estimate business-as-usual (BAU) reductions in NCD mortality risk ($_{40}q_{30}$). For countries not on track to meet SDG 3·4 under BAU, we modelled two intervention scenarios commencing in 2022 that would achieve the target: (1) *prevention* by accelerated reduction in NCD incidence rates; (2) *treatment* by accelerated increases in remission rates and decreases in CFRs.

Australian disease expenditure data, disaggregated by disease phase using New Zealand estimates, were input into a proportional multistate lifetable model to estimate expenditure changes from 2022 to 2040. Assuming similar expenditure patterns across OECD countries, we extended the analysis to estimate future expenditure changes under prevention and treatment pathways to SDG 3·4.

**Findings**

In Australia, current trends project a 25% reduction in NCD 40q30 by 2030, which is short of the 33·3% SDG 3·4 target. Achieving that target requires a 2·53 percentage point (pp) annual acceleration in incidence decline (prevention) or 1·56pp acceleration in CFR reduction and remission increase (treatment). Prevention reduces disease expenditure by up to 0·72% to 3·17% (of US$1310·2 billion under BAU) by 2030 and 2040, respectively; the default treatment scenario, on the other hand, increases disease expenditure by 0·16% in 2022-2030 before reducing disease expenditure by 0·98% in 2031-2040. A treatment scenario that only reduces case fatality increased disease expenditure in the first 9 years, whereas a treatment scenario by only increasing remission rates resulted in savings similar to the prevention scenario.

Among other OECD countries, only Sweden, Ireland, and South Korea (combined sexes) were on track to meet SDG3·4. Prevention and treatment scenarios in other OECD countries showed similar impacts on expenditure to Australia.

**Interpretation**

Whether or not reducing NCD mortality will also save health systems money depends on the pathway taken (prevention or treatment). Nuance is required by researchers, policy makers and advocates if claiming NCD mortality decline will also result in health system savings.

**Funding**

None

**Research in context**

**Evidence before this study**

We searched PubMed for English-language articles published between 1st January 2015 and 24th June 2025, using the terms ("non-communicable diseases" OR "NCDs" OR "chronic disease") AND ((("premature mortality") AND ("Sustainable Development Goals target 3·4" OR "SDG 3·4")) OR ("health expenditure" OR "health savings" OR "economic value")) in the title or abstract. Regarding achieving SDG3·4, this search identified several key studies that have assessed global and regional progress. The NCD Countdown 2030 studies (Lancet 2018, 2020, 2022) revealed that only a small number of high-income countries were on track to meet the SDG 3·4 target, with many experiencing stagnations or increases in NCD mortality. Subsequent analyses (e.g. Martinez et al. 2020; Bray et al. 2021; Feigin et al. 2023; Murthy et al. 2024) have emphasised the rising burden of NCDs such as stroke, cancer, and diabetes, and highlighted persistent health system and policy gaps across both high- and middle-income settings.

Regarding whether reducing NCD mortality saves health system expenditure, economic and modelling studies (e.g. Nugent et al. 2018; Bertram et al. 2018; Devaux et al. 2019; Goryakin et al. 2020) have explored the return on investment of NCD mortality reduction strategies and suggested that preventive measures can yield substantial long-term savings. These studies mostly followed prevention pathways, focusing on the potential benefits of risk reduction and early intervention. To date, there has been limited work comprehensively modelling health system expenditure implications across all OECD countries while accounting for competing risks and long-term outcomes of achieving SDG 3·4 through different intervention pathways.

**Added value of this study**

This study is the first to comprehensively model the health system expenditure impact of preventive and treatment strategies to achieve NCD mortality decline (achieving SDG target 3·4, a one-third reduction in premature NCD mortality from 2015 to 2030) across all OECD countries – allowing for competing morbidity and mortality, and health system expenditure in out years (to 2040) due to people living longer. We used a proportional multistate lifetable model that models incidence rate reduction (i.e. prevention) and case fatality rate reduction with remission rate increase (i.e. treatment strategy) to achieve SDG 3·4 and simulates future changes in disease prevalence and mortality and total health system expenditure. The acceleration in changes in disease rates to achieve SDG3·4 was profound and likely unfeasible for most countries. Nevertheless, a prevention strategy reduces future health system expenditure more than a treatment strategy – and a treatment strategy could increase health system expenditure due to a growing prevalent pool of diseased. A blended prevention-treatment strategy was more feasible in terms of disease rate changes needed and tended to reduce future health system expenditure across all OECD countries. Our study does not assess the actual costs of strategies – just the downstream changes in health system expenditure due to changing population size and disease prevalence, with assumed constant expenditure per person with disease in the future.

**Implications of all available evidence**

While full achievement of SDG 3·4 target is ambitious (especially through prevention), even partial progress can lead to significant health gains and economic savings. Prevention-based strategies, where feasible, reduce health system expenditure more than treatment strategies out to 2040 in OECD countries. With ageing populations, we should consider and optimise strategies to both increase longevity and reduce morbidity and reduce or mitigate impacts on future health system expenditure.

# 1 Introduction

Non-communicable diseases (NCDs) account for 75% of global deaths, killing 43 million people annually, of which 15 million are aged 30-69 years.[1] Health system expenditure in most countries is dominated by NCD treatment.[2] Lead agencies (such as WHO [3]), governments, and researchers[4,5] find that reducing NCD rates is economically advantageous through: reduced health system costs and expenditure; increased workforce productivity; and resultant increased economic growth.

Whilst there is often the view that reducing NCD mortality rates will save health system expenditure[5], this is not necessarily the case. Putting aside the upfront intervention costs, and focusing on downstream changes in health system expenditure driven by prevalent disease, prevention (i.e. reduced disease incidence) should lower health system costs at least initially[6] – but may increase health system expenditure in the medium- to long-term, from people living longer and succumbing to other diseases. Conversely, treatments reducing case fatality rates (without concomitantly increasing remission rates) will increase the prevalent pool of people with disease[7] and increase health system expenditure.[8] An important research question to answer, therefore, is: "What is the impact on future health system expenditure of reducing NCD mortality through prevention or treatment pathways?" Previous studies have not comprehensively examined this research question, yet such information should be useful to society to find pathways to maximise success (reduced or at least deferred health system costs) of success (reduced NCD death rates) rather than failure of success (i.e. lower NCD burden and death rates, but with increasing health system costs).

Previous studies show that achieving NCD risk factor targets (i.e. prevention) can generate substantial but context-dependent savings. For example, Devaux et al estimated that reducing premature NCD4 mortality in France could save €660 million (0·35%) annually.[9] Similarly, Goryakin et al. projected up to €605 billion in savings across the European Union by 2050 under the best-case prevention scenario.[10] But we are not aware of studies examining future health system expenditure impacts of both prevention and treatment pathways to NCD mortality reduction.

Target 3·4 of the Sustainable Development Goals (SDG 3·4) aims to reduce premature mortality from NCDs by one-third (relative to 2015) by 2030,[11] specifically a reduction in the cumulative probability of dying from four NCDs (NCD4-cardiovascular diseases, chronic respiratory diseases, cancers and diabetes) between the ages of 30 and 70 years. Most countries are falling behind in meeting SDG target 3·4.[12,13] An NCD Countdown 2030 report (2018) showed that only five of the 38 OECD nations were on track to achieve a one-third reduction in NCD mortality for both sexes by 2030.[14] While the rate of NCD burden is lowest in high-income countries, with most still decreasing,[14,15] mortality rates have increased in some of these countries.[16] Accelerating the progress towards SDG 3·4 is challenging and demands a comprehensive approach of combining prevention, early detection and treatment strategies across different countries and regions. More specifically, interventions to accelerate mortality decline need to work through at least one of accelerated NCD incidence rate reductions (i.e. prevention) or accelerated decreases in case fatality or increases in remission rates (i.e. treatment). Previous studies neither explicitly model the feasibility of such pathways to SDG 3·4, nor considered what the implications of these alternative pathways might be on future health system expenditure (be that achieving SDG 3·4 specifically, or NCD mortality decline more generally across many countries).[5,9,10,17]

Our study estimates across OECD countries both: 1) the changes in disease rates (incidence for prevention strategy; case fatality and remission rates for treatment strategy) necessary to achieve NCD mortality reduction and reach SDG 3·4; and 2) the impact of these strategies (compared to ongoing business-as-usual disease rate changes) on future health system expenditure from 2022 to 2030 and 2031 to 2040. We focus first on Australia, given it is the primary source of our disease expenditure data,[18,19] but we also extend our analyses to all OECD countries, using previously estimated comparable expenditure by disease across countries.[20] (preprint) Disease expenditure data are inputted to a proportional multistate lifetable model (PMSLT)[21,22] populated with disease rates derived from GBD data, with annual changes in those rates accelerated under prevention and treatment scenarios to achieve SDG 3·4.

## 2 Methods

### 2.1 Data sources and data inputs

We sourced 1990 to 2021 Global Burden of Disease (GBD) disease incidence, prevalence and mortality rates,[23] initial estimates of case fatality rates (CFRs; mortality rate divided by prevalence), and inputted these to the Scalable Health Intervention Evaluation proportional multistate lifetable (SHINE-PMSLT; see Blakely et al.[22] for details) to solve the annual remission rates and generate a smoothed and cohort-coherent dataset of all parameters from 1990 to 2021, for each OECD country.[24]

Australian health system expenditure on the 44 diseases within the NCD4 umbrella grouping was taken from Australian Institute of Health and Welfare estimates,[19] disaggregated by first year of diagnosis, last year of life if dying of the disease, and otherwise prevalent with the disease, using New Zealand data on these relativities.[25] Comparable estimates for all OECD studies were then derived by assuming the relative expenditure by (all – not just the 44 NCD4) disease phase estimate in Australia applied in other OECD countries, with total disease expenditure scaled to an envelope provided by OECD national health accounts[26] and using GBD disease rates[23] (details of method provided elsewhere[20]).

## 2.2 Business-as-usual (BAU) scenario

In the BAU scenario, recent trends in NCD incidence, case fatality and remission rates from 1990 to 2021 were projected to continue until 2030 – thence also giving disease mortality rates to 2030. The risk of dying for the 44 NCDs (**supplementary Table S1**) combined between ages 30 and 70 years ($_{40}q_{30}$) was calculated in each calendar year using a period method, i.e. assuming mortality estimates in each calendar year apply to a 'hypothetical' cohort of people ageing from 30 to 70 years. The percentage reduction in $_{40}q_{30}$ from 2015 to that forecast under BAU in 2030 was estimated for Australia and all other OECD countries, by sex. Further details in Table 1.

## 2.3 Intervention

For sex by country strata where the forecast business-as-usual (BAU) reduction in $_{40}q_{30}$ from 2015 to 2030 was less than a third (i.e. the SDG 3·4 was not met), we modelled two main intervention scenarios such that the forecast $_{40}q_{30}$ in 2030 was one-third less than in 2015: (1) a prevention scenario through the same percentage point acceleration from 2021 in the BAU annual percentage change in incidence rates for all sex by age by 44 NCD groupings; (2) a treatment scenario through the same percentage point acceleration from 2021 in both remission rates and CFRs. Further details in **supplementary Table S2**. It seems unlikely that, in reality, one can focus accelerated changes in disease rates only to 30- to 70-year-olds, without also affecting 70+ year olds. Therefore, we assumed that the accelerations applied to all ages, which impacts the disease expenditure calculations.

## 2.4 Data analysis

The PMSLT was run with expected or median values of all input values (i.e. Monte Carlo analyses with input parameter uncertainty were not included in this study) in a closed cohort. A time horizon to 2040 was used, with results presented up to 2030 and for 2031 to 2040. For the 2040-time horizon, each of the three intervention scenarios from 2030 returned to their BAU annual percentage changes (APCs) in incidence, remission and CFRs.

Expenditure estimates were consumer price index-adjusted to the 2021 base year, and purchase power parity adjusted to 2019 USD. Results in the paper are undiscounted. The main analyses were conducted for Australia and estimated percentage changes in total health system expenditure and expenditure per person years (to adjust for the increasing population size from NCD reduction) among 30+ years compared to BAU for prevention and treatment scenarios. Further details in supplementary file (*p2*).

Supplementary analyses estimated health system expenditure for a blended intervention scenario with approximately half the acceleration in incidence rate changes in the prevention scenario and half the acceleration of remission and CFR changes of the treatment scenario. We also include a treatment by CFR-only scenario and a treatment by remission-only scenario.

## 3 Results

### 3.1 Additional APCs need to achieve SDG 3·4

For disease rate trends in Australia observed to 2021, extended out to 2030 (i.e. BAU forecast), the period risk of NCD death from ages 30 to 70 ($_{40}q_{30}$) in 2030 is 24·5% and 24·3% less than in 2015 for women and men, respectively. To achieve the 33·3% SDG 3·4 target reduction in $_{40}q_{30}$ from 2015 to 2030 for women, under the prevention scenario, a 2·53 percentage point (pp) annual acceleration of incidence rate reductions is required for all 44 NCD4 diseases (**Supplementary Table S3**). Expressed another way, each of the NCD4 diseases needs to have a disease incidence rate in 2030 that is 20·4% (1-exp [-(2030-2021) × 0·0253]) less than forecast under BAU. For women to achieve SDG 3·4 under a treatment scenario, a 1·41pp acceleration in both case fatality and remission rates is required (or CFRs 11·9% less and remission rates 13·5% greater than BAU forecasts in 2030). For Australian men, a 2·51pp annual acceleration of incidence rate decreases is required under the prevention scenario, or a 1·56pp acceleration in both case-fatality and remission rate trends is required under the treatment scenario.

Across OECD countries, only Sweden, Ireland, South Korea, and Switzerland for men, and Israel and Ireland for women, will achieve SDG 3·4 under BAU (**Figure 1**). Combining both sexes, Sweden, Ireland and South Korea were estimated to meet SDG 3·4. We estimated the additional absolute percentage change from the BAU APCs that is required to achieve SDG 3·4 for countries that failed to meet the target under the BAU scenario. On average across OECD countries, a 3·6 and 4·0pp additional annual decline in incidence rates for men and women, respectively, was required under the prevention scenario, with the largest acceleration required in Mexico among both men (16·8pp) and women (11·8pp) (**supplementary Table S3**). In the treatment scenario, the APCs for both CFRs and remission rates were increased on average by 2·2pp for both men and women, with the highest additional APC rate of 6·1% for men in Mexico and 4·4% for women in Iceland (**supplementary Table S3**).

The age-standardised disease prevalence in 2030 for ischemic heart disease (IHD), stroke, diabetes and colorectal cancer decreased on average by about 9·4% for the prevention scenario compared to BAU and increased by about 1·3% for the treatment scenario (**Supplementary Figure S1**).

## 3.2 Changes in health system disease expenditure for 30+ year olds for prevention and treatment scenarios to achieve SDG 3·4

For Australia BAU health system disease expenditure is forecast to be US$1133 billion in 2022-30 and US$1310 billion in 2031-40 (**Table 2**). The prevention scenario reduces this expenditure by 0·72% and 3·17%, respectively, whereas the default treatment scenario (acceleration in both case fatality rate reduction and remission rate increases) increases disease expenditure by 0·16% in 2022-30 before reducing disease expenditure (i.e. savings) by 0·98% in 2031-40. A treatment scenario by only reducing CFRs costs 0·30% more than BAU in 2022-30, although it does have modest savings in 2031-40 of 0·56%. Treatment by only increasing disease remission rates (i.e. 'cure') has savings similar to the prevention scenario. The blended scenario (i.e. equal contributions of both prevention and default treatment) has – unsurprisingly – impacts on health system expenditure between prevention-only and default treatment-only scenarios.

**Figure 2** shows the projected impact of the prevention and default treatment scenarios on health expenditure across all 34 OECD countries from 2022 to 2030. Similar to Australia, the pattern across OECD countries is of the prevention scenario (yellow bars) resulting in savings in 2022-30 (average 0·8%, range 0·1% for Switzerland to 2·1% for the USA) that increase further in 2031-40 (average 3·3%, range 1·8% for Switzerland to 6·8% for the USA). The default treatment scenario (blue bars) resulting in increased expenditure in 2022-30 (average increased expenditure 0·3%, 0·1% in multiple countries to 0·9% for Hungary), switching to savings in most countries for 2031-40 (average 0·6% savings, range -1·7% for Hungary (i.e. increased expenditure) to 1·9% saving for Slovenia). Scenario results for all OECD countries for blended prevention and treatment, treatment by decreased case fatality only, and treatment by increased remission only are shown in **Supplementary Figure S2.**

### 3.3  Health system disease expenditure as person-year rates

Prevention and treatment scenarios change the number of people alive in the future – so it is important to also consider changes in disease expenditure per person years lived, as shown in **Figure 3** for Australia. Using all ages in person year rates for BAU and intervention scenarios (**Figure 3b**) sees percentage savings all modestly increase (and increased expenditure in some treatment scenarios decreases) compared to raw total expenditure percentage changes in **Figure 3a**, i.e. factoring in a large population size makes results more favourable. **Figure 3c** uses person years aged 25 to 64 years as the denominator, to represent the impact from the perspective of changes to the working age population, with all rates now intermediary between those in **Figure 3a** and **Figure 3b** given that reducing NCD mortality (be it through prevention or treatment) does increase the number of working age people – but not as much in relative terms as it increases the population aged 65+ years. **Figure 3d** takes the concept of working age adults one step further, to include people aged some days to weeks older than exactly 65 years of age – where that age extension is to the exact age under the prevention or treatment scenario that has the same morbidity rate of an exactly 65-year-old in BAU. Put another way, people (on average) are assumed to work longer pro-rata, deferring the morbidity into the future, given reduced NCD prevalence. The percentage changes for this extended working age population (**Figure 3d**) improve modestly compared to just using 25-64-year-old person years as the denominator (**Figure 3c**).

Percentage changes in expenditure rates for all-age person-year and 25-64-year-old person-year denominators, for all OECD countries, are shown in **Supplementary Figure S2** and **Supplementary Figure S3**; the pattern is similar to that for Australia. Supplementary **Figure S4** shows health system disease expenditure for all OECD countries under the blended, treatment by CFR only and treatment by remission only scenarios.

## 4  Discussion

As with other studies,[12,27,28] we also find that most OECD countries are not on track to achieve a third reduction in premature NCD mortality risk among 30-70-year-olds from 2015 to 2030. In Australia, achieving SDG 3·4 under a prevention scenario required a 2·5pp greater annual decline in incidence rates than under BAU. Under the treatment scenario, a 1·5pp acceleration in both CFR decline and remission increase was needed. Across the OECD countries, these values averaged 3·8pp for prevention and 2·2 pp for treatment. Generally, a prevention strategy to achieve NCD mortality rate decline saved health system expenditure, whereas a treatment strategy was mixed and modest in its impacts due to prevention reducing disease prevalence, but treatment not necessarily doing so.

The savings under prevention were due to ambitious – unlikely feasible – annual declines in NCD incidence rates. However, negligible short-run cost savings (if not increased expenditure) under the treatment scenario demonstrate that reducing NCD mortality does not necessarily lead to health system savings, due to the combined effects of modest, if any, changes in disease prevalence (treatment scenario) and more people alive for longer developing other diseases (both prevention and treatment scenarios).

This study illustrates the need for a nuanced and more sophisticated analysis of how policies we enact now to alter NCD rates impact health system expenditure in the future. First, we need to tally up not only NCD disease expenditure, but all disease expenditure, to estimate the full impact on future health system expenditure. Second, whilst considering just changes in gross expenditure (Figures 2a and 3a) is a good starting point, we should really incorporate the gains in population size under intervention policies (especially prevention), which spreads future health system expenditure across a larger number of people (Figures 2b and 3b). Third, we need to consider who is funding future health services to estimate impacts of sustainability and expenditure 'rates' in the future: Figures 2c and 3c use 25-64 year olds as a proxy for taxpayers funding health systems, and Figures 2d and 3d take the analysis further to consider how the age of 65 is not some 'fixed' dependency threshold ratio but is amenable to change – including change from a population with less morbidity being able to work longer (noting that there are many other considerations of individuals and society as to how long people work for). Regardless of the perspective, prevention (if incidence rate reductions are feasible and achievable) appears to be a better strategy than treatment.

It is one thing to demonstrate that prevention strategies, if implemented at ambitious levels of incidence decline, could save health system costs and reduce mortality. It is quite another to confront the blunt reality that achieving such declines is politically and practically challenging. Modifiable risk factors like tobacco use, diet, and physical inactivity are deeply embedded in social and economic systems that resist rapid change. Even in well-resourced OECD countries, preventive interventions rarely deliver the scale of impact needed, and without decisive political will and innovation, relying on prevention risks overpromising, while health systems must prepare for sustained treatment demand.

**Limitations**

Our study has many limitations and assumptions, some of which are also opportunities for further research and policy focus. First, we assumed that the cost of someone with a given disease in the future is the same as today. These disease-specific costs will likely change. But whether they increase or decrease is moot, and amenable to policy interventions themselves. For example, costs may decrease if health system delivery is more efficient and productive, treatments are cheaper, and (over and above what we modelled in this paper) treatments also reduce disease severity (and thence some treatment costs) in addition to 'just' changing disease case fatality and remission rates. On the other hand, if prioritisation of cost-effective treatment options is poorly implemented, and citizen expectations for more (expensive) treatment increase, then disease-related expenditure will inexorably increase, perhaps outweighing any gains from much accelerated disease incidence reduction. Third, our study uses a comparable set of disease expenditure estimates across all OECD countries, which is a strength for 'big picture' analyses. But the estimates are reflective of relativities in disease expenditure estimated in Australia, extrapolated to other countries; there is a need for improved disease expenditure estimates over multiple countries. Unless the 'true' differences in expenditure across countries are strongly correlated with which NCDs are going to be prominent or not in the future, the patterns observed in this study are unlikely to change much.

The SDG 3·4 indicator relied on the International Classification of Diseases and Related Health Problems, 10th Revision (ICD-10), to classify conditions within the four NCD disease groups, which differs slightly from the categorisation used by the GBD.[23] As a result, certain rare diseases classified under "other conditions" by GBD (such as some cardiovascular and circulatory diseases) were excluded from this analysis, although they are included in the SDG 3·4 indicator due to differences in disease coding systems. However, the approach adopted here uses the most closely aligned classification available, and most of the excluded diseases are rare, contributing only a small fraction to overall NCD-related mortality. Therefore, this limitation is unlikely to have a significant impact on the results.

The PMSLT model assigns morbidity and associated expenditures independently to each disease in this analysis, potentially overlooking compounded costs from comorbidities. When patients suffer from multiple conditions, total medical costs often exceed the sum of the individual costs for each condition,[22,25] possibly suggesting that savings in expenditure are slightly underestimated. However, the initial expenditure estimates were adjusted for comorbidities,[20] which helped mitigate this underestimation.

Our study does not estimate the cost-effectiveness of prevention and treatment strategies. However, if readers have intervention strategies in mind with estimable accelerations in disease rates under BAU, and an estimated cost for such an intervention, then it is but a small step to estimate the net strategy cost impacts over a 20-year timeline (i.e. intervention cost minus cost savings), and a step further to estimate health gains (e.g. health adjusted life years) to estimate cost-effectiveness.

Finally, our analysis did not account for population changes due to births or migration, which may influence future disease burden and health expenditures, particularly over longer time horizons.

**Conclusion**

This study offers new insights into the economic implications of achieving SDG 3·4 in OECD countries and the different impacts of strategies to achieve NCD mortality decline on future health system expenditure. The results also highlight the effects of extended longevity and competing morbidity. There is a need for more nuanced and sophisticated thinking and quantification of prevention and treatment strategies on future health system expenditure, and opportunities with such analysis and arising policy actions to assist in achieving more financially sustainable health systems in the future.

**Acknowledgement**

This research received no specific grant from any funding agency in the public, commercial, or not-for-profit sectors. We are grateful to Joe Dieleman and Phillip Clarke for their insightful comments and suggestions on this manuscript. Their contributions have helped refine the presentation of our work.

**Declaration of interests**

Jingjing Yang is currently employed by KeAi Communications Co., Ltd. All other authors declare no conflicts of interest.

**Data sharing**

All disease burden data used in the modelling are freely accessible via the Global Burden of Disease (GBD) website. Disease-specific health expenditure data are available from the Australian Institute of Health and Welfare.

# Tables and figures

**Table 1. Business-as-usual (BAU) parameters**

| Parameter | Data source | Trend and scenario analysis |
|---|---|---|
| Population | GBD [23] | Baseline population counts by age and sex in 2021. |
| All-cause mortality rates | GBD [23] | Forecast all-cause mortality rate by sex and age to 2030, then hold constant |
| Disease-specific inputs of incidence, prevalence, mortality, remission, and case fatality rates (CFRs) | GBD [23] | Used data from 1990 to 2021 to estimate incidence, CFR and remission rates by sex and age cohort, using a series of algorithms described elsewhere.[29] These were forecast forward to 2030 for a 10-year time horizon and forecasted to 2040 for a 20-year time horizon. |
| Disease-related health expenditure | Australian Institute of Health and Welfare (AIHW) [18,19] | Disease-related health expenditure by sex and age sourced from the AIHW was further stratified by disease-phase – as first year of diagnosis, last year of life if dying of that disease, and otherwise prevalent–using New Zealand estimates of relative variation by phase. [25] OECD-specific estimates were generated by scaling to total health expenditure (below), given disease incident cases, prevalent cases and cases dying of that disease (see text). |
| Total health system expenditure | OECD national health accounts [26] | Expenditure by area of expenditure for each OECD country |

**Table 2. Changes compared to BAU in health system expenditure in Australia for various preventive and treatment interventions that achieve SDG3.4**

|  | 2022-30 | | 2031-40 | |
|---|---|---|---|---|
|  | USD billions† | % savings c.f. BAU | USD billions† | % savings c.f. BAU |
| **Business-as-usual (BAU)** | $1133·28 | *NA* | $1310·27 | *NA* |
| **Intervention scenarios- savings in health expenditure <u>compared</u> to BAU** | | | | |
| **Prevention** | | | | |
| Incidence rate changes only | $8·15 | 0·72% | $41·49 | 3·17% |
| **Treatment** | | | | |
| Default: Equal acceleration CFR and remission rates | -$1·82 | -0·16% | $12·80 | 0·98% |
| CFR acceleration only | -$3·35 | -0·30% | $7·34 | 0·56% |
| Remission rate acceleration only | $7·20 | 0·64% | $42·67 | 3·26% |
| **Blended** | | | | |
| Even contributions for prevention and 'default' treatment | $3·29 | 0·29% | $27·66 | 2·11% |

†In 2019 real dollars; c.f.-compared to

Note. Prevention – accelerated annual percentage change in incidence rate reduction from 2021 to 2030 then BAU annual percentage change trends until 2040; Treatment – Accelerated annual percentage change in case fatality reduction and remission increase from 2021 to 2030 then BAU trends until 2040; Blended – a blended scenario with approximately half the acceleration in incidence rate changes in the prevention scenario and half the acceleration of remission and CFR changes of the treatment scenario; Treatment by CFR – Accelerated annual percentage change in case fatality reduction (without accounting for remission) from 2021 to 2030 then BAU annual percentage change trends until 2040; Treatment by remission – Accelerated annual percentage change in remission rate increase (without accounting for case fatality rate) from 2021 to 2030 then BAU annual percentage change trends until 2040.

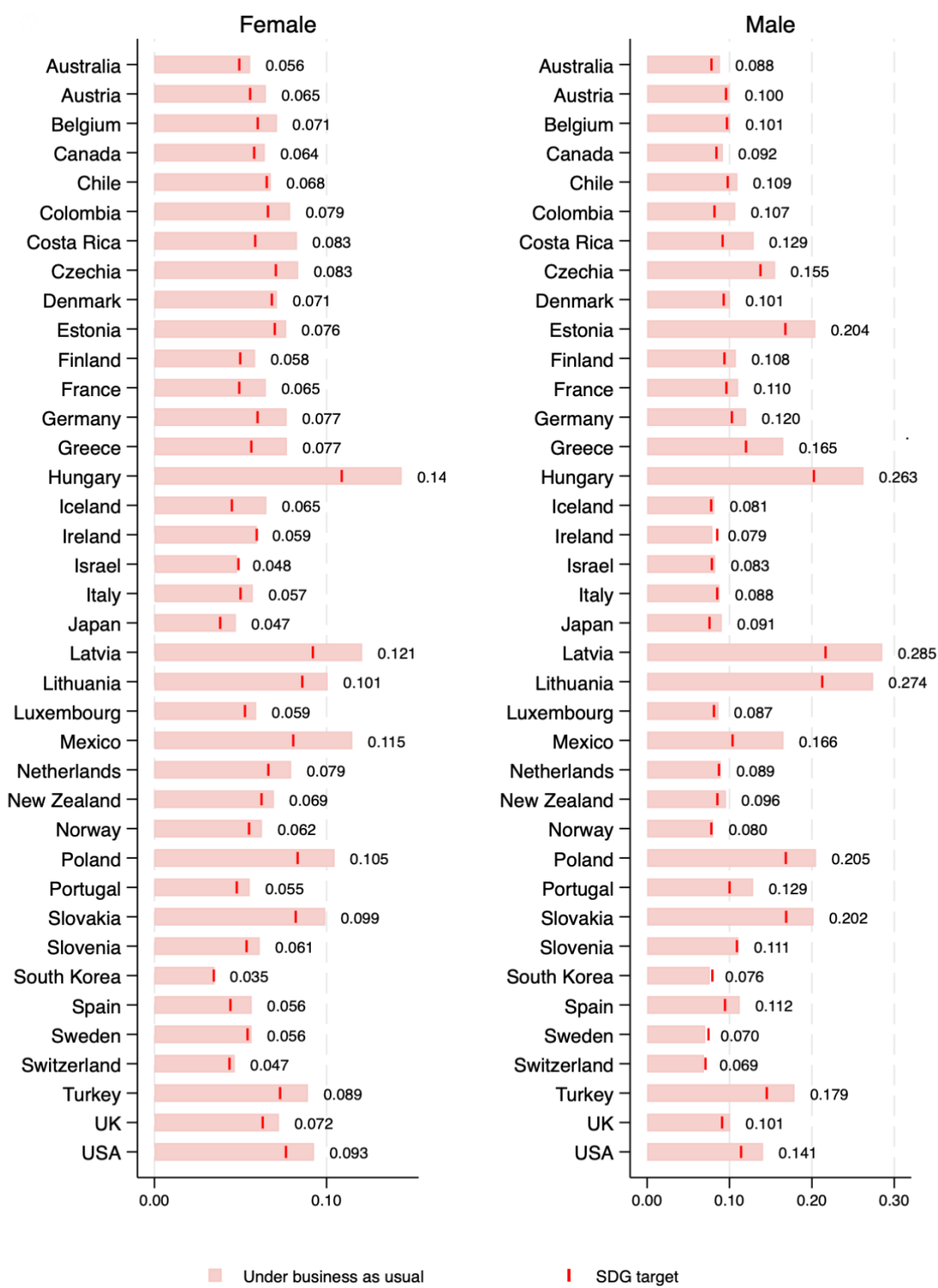

**Figure 1. Mortality risk from age 30 to 70 ($_{40}q_{30}$) in 2030 across 38 OECD countries, separately by sex**
Footnote: **Supplementary Table S3** has data used in the construction of Figure 1.

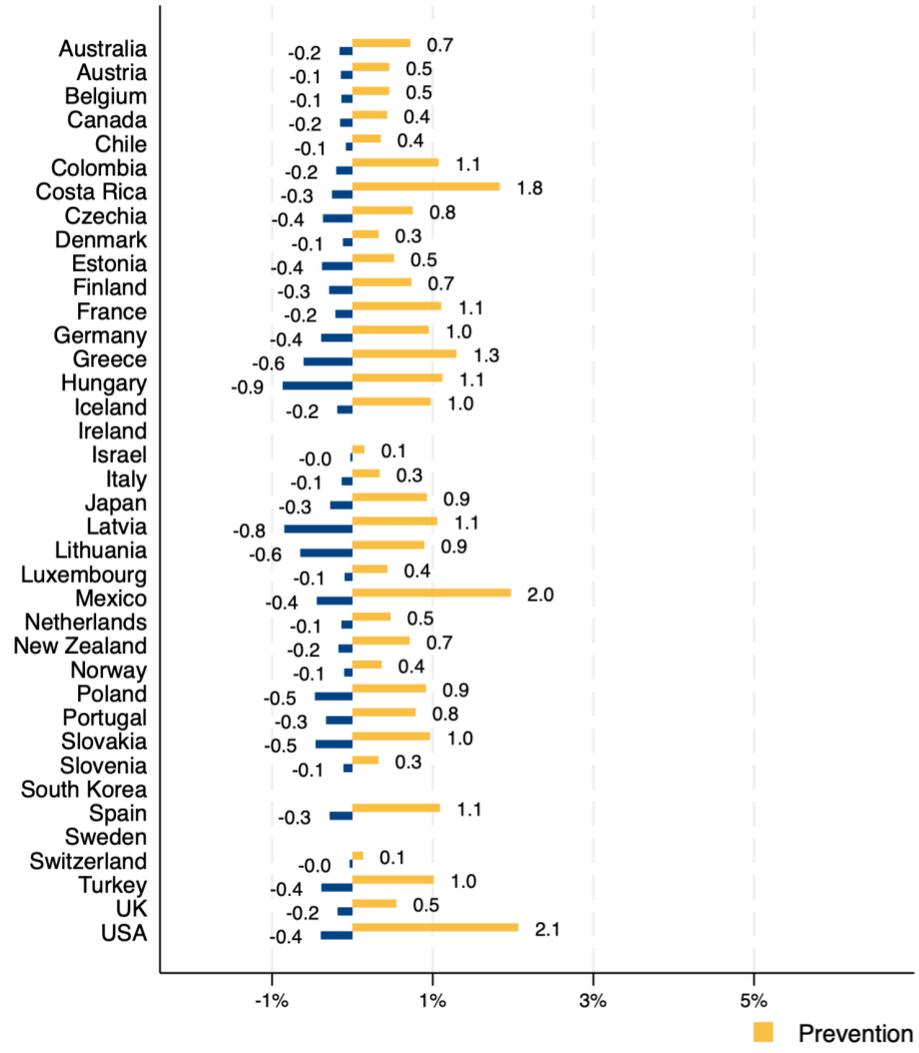 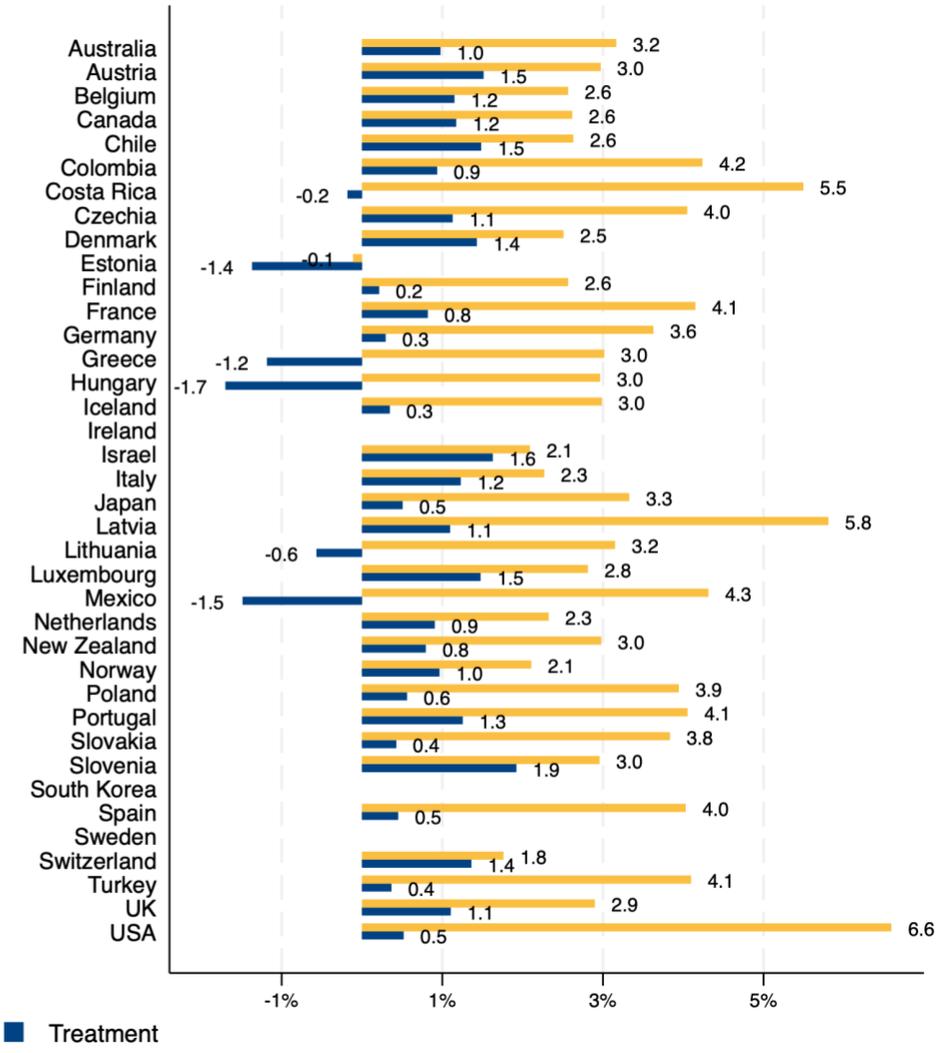

a. 2022-2030     b. 2031-2040

**Figure 2. Savings in health system expenditure (as percentage of BAU) by OECD country for 30+ year olds from (a) 2022-2030 and (b) 2031-2040 for prevention and treatment scenarios that achieve SDG 3·4, undiscounted, sexes combined.**

Footnote. Prevention – accelerated annual percentage change in incidence rate reduction from 2021 to 2030, then BAU annual percentage change trends until 2040; Treatment – Accelerated annual percentage change in case fatality reduction and remission increase from 2021 to 2030, then BAU annual percentage change trends until 2040. Ireland, South Korea, and Sweden are not modelled as they are likely to meet SDG 3·4 in 2030 under the business-as-usual scenario

2022-2030　　　　　　　　　　　2031-2040

a. Total health expenditure

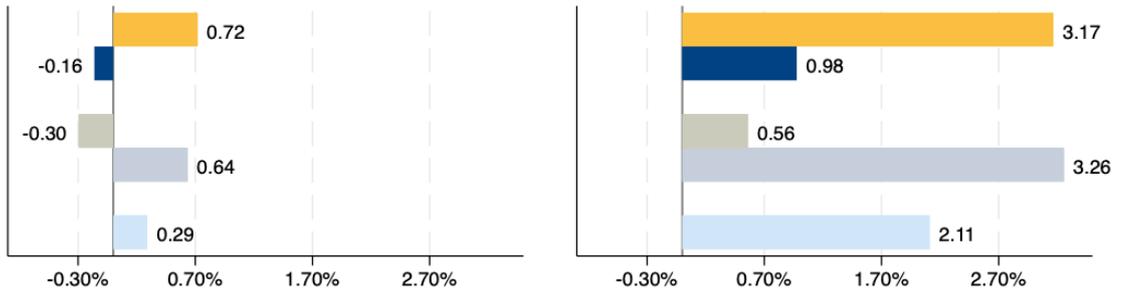

b. Health expenditure per person years

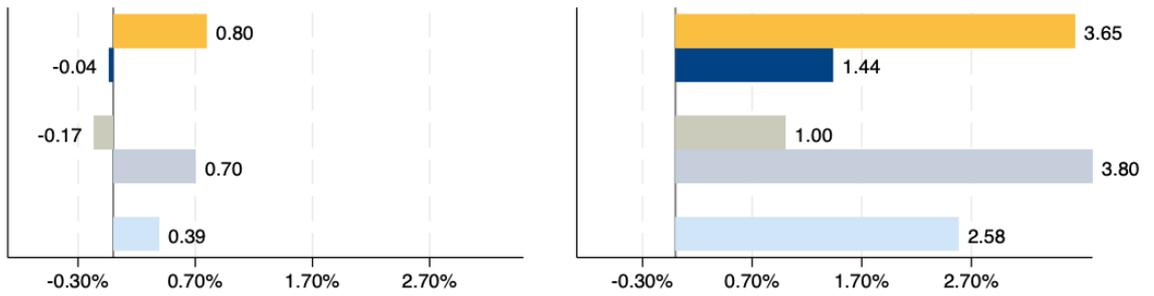

c. Health expenditure per person years aged 25-64

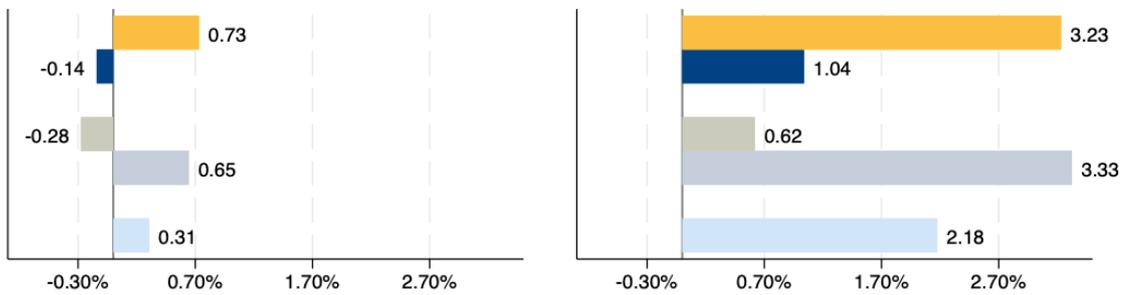

d. Health expenditure per person years aged 25 to age with equivalent morbidity to that in BAU for 65 year olds

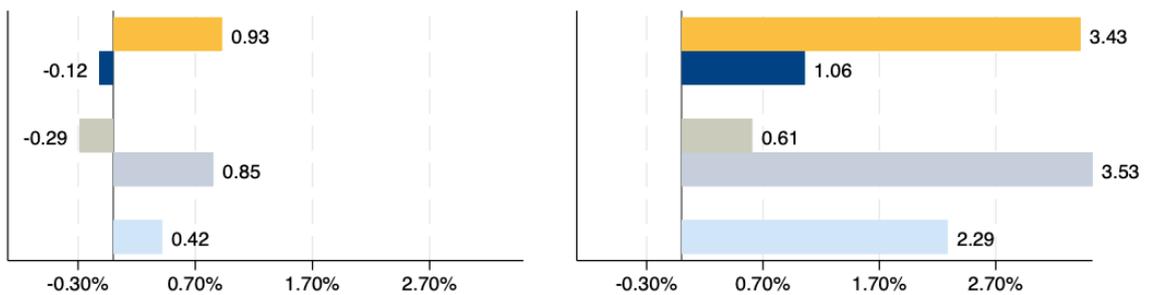

■ Prevention　■ Treatment　■ Treatment by CFR only　■ Treatment by Remission only　■ Blended

**Figure 3. Savings in health system disease expenditure in Australia, 2022-2030 (left) and 2031-2040 (right) for different prevention and treatment scenarios that achieve SDG 3·4, undiscounted, sexes combined, expressed as percentage changes from BAU in: a) raw total health expenditure; b) expenditure rate per person years all ages; c) expenditure rates per person years aged 25-64; and d) expenditure rate per person years aged 25 to age with morbidity equivalent to 65 year olds in BAU.**

Footnote. Prevention – accelerated annual percentage change in incidence rate reduction from 2021 to 2030 then BAU annual percentage change trends until 2040; Treatment – Accelerated annual percentage change in case fatality reduction and remission increase from 2021 to 2030 then BAU trends until 2040; Blended – a blended scenario with approximately half the acceleration in incidence rate changes in the prevention scenario and half the acceleration of remission and CFR changes of the treatment scenario from 2021 to 2030 then BAU annual percentage change trends until 2040; Treatment by CFR – Accelerated annual percentage change in case fatality reduction (without accounting for remission) from 2021 to 2030 then BAU annual percentage change trends until 2040; Treatment by remission – Accelerated annual percentage change in remission rate increase (without accounting for case fatality rate) from 2021 to 2030 then BAU annual percentage change trends until 2040.